\documentclass[aip,reprint]{revtex4-1}
\usepackage[T1]{fontenc}
\usepackage[utf8]{inputenc}
\usepackage{graphicx}
\usepackage{amsmath}
\usepackage{amssymb}
\usepackage{color} 
\usepackage{pifont}

\newcommand{\beq}{\begin{equation}}
\newcommand{\eeq}{\end{equation}}

\begin{document}
\title{Intermolecular Cross-Correlations in the Dielectric Response of Glycerol}

\author{Jan Philipp Gabriel$^{\text{1,2,}\hspace{-2pt}}$}

\email{Jan.Gabriel@fkp.physik.tu-darmstadt.de}
\author{Parvaneh Zourchang$^{\text{1}}$}
\noaffiliation
\author{Florian Pabst$^{\text{1}}$}
\noaffiliation
\author{Andreas Helbling$^{\text{1}}$}
\noaffiliation
\author{Peter Weigl$^{\text{1}}$}
\noaffiliation
\author{Till Böhmer$^{\text{1}}$}
\noaffiliation
\author{Thomas Blochowicz$^{\text{1,}\hspace{-1pt}}$}
\email{thomas.blochowicz@physik.tu-darmstadt.de}
\noaffiliation

\affiliation{Institut für Festkörperphysik, Technische Universität Darmstadt, 64289 Darmstadt, Germany.}

\affiliation{Arizona State University, School of Molecular Sciences, Tempe, AZ 85281, United States}

\date{\today} 

\begin{abstract}
We suggest a way to disentangle self- from cross-correlation contributions in the dielectric spectra of glycerol. Recently it was demonstrated for monohydroxy alcohols that a detailed comparison of the dynamic susceptibilities of photon correlation and broadband dielectric spectroscopy allows to unambiguously disentangle a collective relaxation mode known as the Debye process, which could arises due to supramolecular structures, and the $\alpha$-relaxation, which proves to be identical in both methods. In the present paper, we apply the same idea and analysis to the paradigmatic glass former glycerol. For that purpose we present new light scattering data from photon correlation spectroscopy measurements and combine these with literature data to obtain a data set covering a dynamic range from $10^{-4}-10^{13}\,$Hz. Then we apply the above mentioned analysis by comparing this data set with a corresponding set of broadband dielectric data. Our finding is that even in a polyalcohol self- and cross-correlation contributions can approximately be disentangled in that way and that the emerging picture is very similar to that in monohydroxy alcohols. This is further supported by comparing the data with fast field cycling NMR measurements and dynamic shear relaxation data from the literature, and it turns out that, within the described approach, the $\alpha$-process appears very similar in all methods, while the pronounced differences observed in the spectral density are due to a different expression of the slow collective relaxational contribution. In the dielectric spectra the strength of this peak is reasonably well estimated by the Kirkwood correlation factor, which supports the view that it arises due to dynamic cross-correlations, which were previously often assumed to be negligible in dielectric measurements.
\end{abstract}

\maketitle 


\section{Introduction}
In the study of supercooled liquids, glycerol has a prominent position as one of the pet systems, probably due to its very low tendency to crystallize and also because of its large dipole moment, which make it a good candidate particularly for dielectric investigations \cite{Davidson:1951,Paluch:1996,Schneider:1998,Ryabov:2003,Hensel:2004,Pronin:2010}. But in fact, glycerol has been studied intensively with a great number of experimental techniques to answer fundamental questions of glass and fluid dynamics, including, besides dielectric spectroscopy, nuclear magnetic resonance,\cite{Doess:2001,Gainaru:2008,Flaemig:2019} dynamic light scattering,\cite{Brodin:2005a,Vogel:2001} neutron scattering \cite{Wuttke:1994,Towey:2011,Gupta:2015} and many other techniques.\cite{Beevers:1980,Yee:1996,Jensen:2018,Klieber:2013,Bo:2016,Kremer:2018} In terms of possible interactions, however, glycerol is by no means a ``simple'' liquid, and thus its status as a ``fruit fly'' of glass research has recently been questioned.\cite{Niss:2018} And indeed, glycerol as a polyalcohol is more complicated than a typical van der Waals liquid, because of three  OH-groups, which are able to form H-bonds in the liquid. At first glance, the spectral shape of a dynamic susceptibility in polyalcohols is hardly different from that of a van der Waals liquid: For example the dielectric loss shows a primary process which is usually identified as the collective structural or $\alpha$-relaxation \cite{Doess:2001}, which is usually thought to be related to the viscosity. However, recently a dynamic shear study revealed an additional relaxation process in glycerol, which is slower than the $\alpha$-relaxation.\cite{Jensen:2018} This is a remarkable observation regarding the fact that only one single peak is observed in dielectric spectroscopy.

In terms of complexity, monohydroxy alcohols with only one OH-group per molecule should in some sense be in between van der Waals liquids and polyalcohols. Many investigations were carried out on monohydroxy alcohols, to single out the influence of H-bonds in this seemingly simple situation.\cite{Boehmer:2014a}  In contrast to van der Waals liquids and polyalcohols many monohydroxy alcohols show a clear two-peak structure in the dielectric loss. At lower frequencies than the $\alpha$-relaxation, a peak with mostly a single-exponential or Debye-like relaxation shape is observed in the spectrum. This peak is attributed to the dynamics of transiently hydrogen-bonded supramolecular chains.\cite{Gainaru:2010a} As the permanent dipole moments are preferentially oriented along the H-bonded structures, the dielectric Debye process mainly reflects intermolecular cross-correlations. Accordingly, in many monohydroxy alcohols the relaxation strength is significantly larger than what is expected from uncorrelated single-molecule dipole moments and the amount is quantified by the Kirkwood correlation factor.\cite{Kirkwood:1939a} In previous investigations of monohydroxy alcohols it was demonstrated that self- and cross-correlation contributions can well be separated experimentally by comparing photon correlation with dielectric data, where in the majority of cases the former turns out to reflect the self-correlation part of the spectrum\,\cite{Gabriel:2018,Gabriel:2018a,Gabriel:2017a,Boehmer:2019a,Weigl:2019}. We note that a similar analysis was also successfully applied to a diol (ethylene glycol), in the pioneering work of Fukasawa et al.~\cite{Fukasawa:2005}, who were able to distinguish the above mentioned contribtions in the high frequency regime, using dielectric and Raman spectroscopy.

Assuming that monohydroxy alcohols are systems with intermediate complexity between van-der-Waals liquids and complex polyalcohols, the question arises whether polyalcohols really show a spectrum that is qualitatively different from monohydroxy alcohols in that it does not show a cross-correlation contribution in the dielectric spectrum or whether such a contribution is simply merged with the $\alpha$-relaxation. The two peak structure recently observed in shear mechanical data of glycerol \cite{Jensen:2018} and also the results obtained in ethylene glycol \cite{Fukasawa:2005} might be a hint that the latter is the case in polyalcohols and that self- and cross-correlation contributions are simply harder to disentangle than in monohydroxy alcohols. Conceptionally one might imagine a less directed total dipole moment in the network structure of polyalcohols than in the linear chain structure of monohydroxy alcohols.

In the present work our approach will be to apply a procedure similar to the one used for the analysis of the dielectric loss in monohydroxy alcohols \cite{Gabriel:2017a,Gabriel:2018,Gabriel:2018a,Boehmer:2019a}. For that purpose we present a broadband light scattering data set of glycerol, where our own photon correlation data are merged with Tandem Fabry-Perot and Raman data from the literature to cover a dynamic range from $10^{-4}-10^{13}\,$Hz. By comparison with broadband dielectric data we suggest a way to disentangle self- and cross-correlation contributions in the dielectric data. Further support for this approach is derived  from the discussion of nuclear magnetic resonance (NMR) fast field cycling measurements and dynamic shear data, both from the literature.


\section{Experimental background}
Samples of glycerol (Karl Roth, >99.7\%),  were used from a new bottle and filtered into a photon correlation spectroscopy (PCS) sample-cell by using a 200\,nm hydrophilic syringe filter, while the dielectric samples were prepared from the same bottle without further purification. Everything was prepared as quickly as possible to minimize unwanted water contamination.

In terms of temperature measurements all temperature environments were carefully calibrated to achieve an overall accuracy of $\pm 0.5\,$K. Broadband dielectric spectroscopy (BDS) was performed using a Novocontrol Alpha-N High Resolution Dielectric Analyzer in combination with a time domain dielectric setup, reported in detail in Ref.~\citenum{Rivera:2004a} and an Agilent E49991A Impedance Analyzer. Combined data sets of the complex dielectric susceptibility $\varepsilon^*(\omega)$ were obtained, covering the frequency range of $10^{-4} - 10^8\,$Hz.

Light scattering experiments were performed in vertical-horizontal (VH) depolarized geometry. In the high frequency range a Tandem Fabry-Perot interferometer was used (TFPI, J.R. Sandercock, $3\cdot10^8-10^{12}$\,Hz) for a few reference measurements to ensure matching temperatures and correlation times between literature data and the data of our own laboratory. The PCS experiments were performed under a scattering angle of $90^\circ$ in a setup already described earlier in detail.\cite{Blochowicz:2013a,Gabriel:2015b,Pabst:2017a} The measured intensity autocorrelation function $g_2(t)=\langle I_{\text{s}}(t)I_{\text{s}}(0)\rangle /\langle I_{\text{s}}\rangle ^2$ was converted into the electric field autocorrelation function $g_1(t)=\langle E^*_{\text{s}}(0)E_{\text{s}}(t)\rangle/\langle \vert E_{\text{s}}\vert\rangle^2$ by using the Siegert relation for partial heterodyning as described in more detail by Pabst et al.\,\cite{Pabst:2017a}\footnote{The required constant $A_\text{fast}$, describing the strength of fast molecular dynamics not captured by the used correlators and not removed from the spectrum by the used $2\,\text{nm}$-filter, was estimated as $0.05$ for glycerol on the basis of the TFPI measurements.}.

In order to allow for a direct comparison of BDS and light scattering data, both data sets are used in the same representation in the frequency domain, i.e. the dielectric loss  $\varepsilon''(\omega)$ is compared with the imaginary part of the complex light scattering susceptibility $\chi''(\omega)$, which is calculated from $g_1(t)$ through Fourier transformation:\cite{Kremer:2002a}
\begin{equation}
\chi''(\omega)\propto \omega\int^\infty_0\, g_1(t)\cos(\omega t)\,dt
\end{equation}
By using the Filon rule of integration\cite{Filon:1929a} the transformation can be carried out for discrete datasets on logarithmic scales, resulting in a spectral representation of the depolarized dynamic light scattering data as generalized susceptibility $\chi''(\omega)$. Because PCS itself is not able to determine absolute relaxation strengths, $\chi''(\omega)$ needs to be renormalized after Fourier transformation. This can be achieved at least to reasonable approximation by combining these data with TFPI measurements, as shown further below. The latter are either normalized to a temperature dependent absolute scattering intensity or, as e.g.\ applied by Brodin et al.,\cite{Brodin:2005a} by using a particular Raman line.

When comparing BDS and light scattering susceptibilities, one has to keep in mind that, although both reflect molecular reorientation, BDS observes a correlation function of a vectorial quantity, the molecular dipole moment, while the PCS observable is connected with a tensorial quantity, the anisotropic molecular polarizability.\cite{Kremer:2002a,Berne:1976a} This leads to reorientational correlation functions that are expressed in terms of Legendre polynomials of rank $\ell$ as:
\begin{equation}
\Phi_\ell(t)=\langle P_\ell(\cos\theta(t))\rangle,
\label{eq:legendre}
\end{equation}
with $\ell=1$ for BDS and $\ell=2$ for light scattering, with $\theta(t)$ being the angle between the positions of the respective molecular axis at times $0$ and $t$. If the optical anisotropy and the permanent dipole moment are mainly located at the same molecular entity and the dynamic processes under consideration are isotropic to good approximation, then relations between $\Phi_1$ and $\Phi_2$ can be established depending on the geometry of the underlying process. In case of random angle jump of the molecules the correlation function becomes independent of $\ell$,\cite{Berne:1976a} and under certain conditions, also small angle based reorientation geometries can lead to approximately $\ell$-independent correlation functions.\cite{Diezemann:1998}


\section{Results and data analysis}
\begin{figure}[t]
\centering
\includegraphics[width=8.8cm]{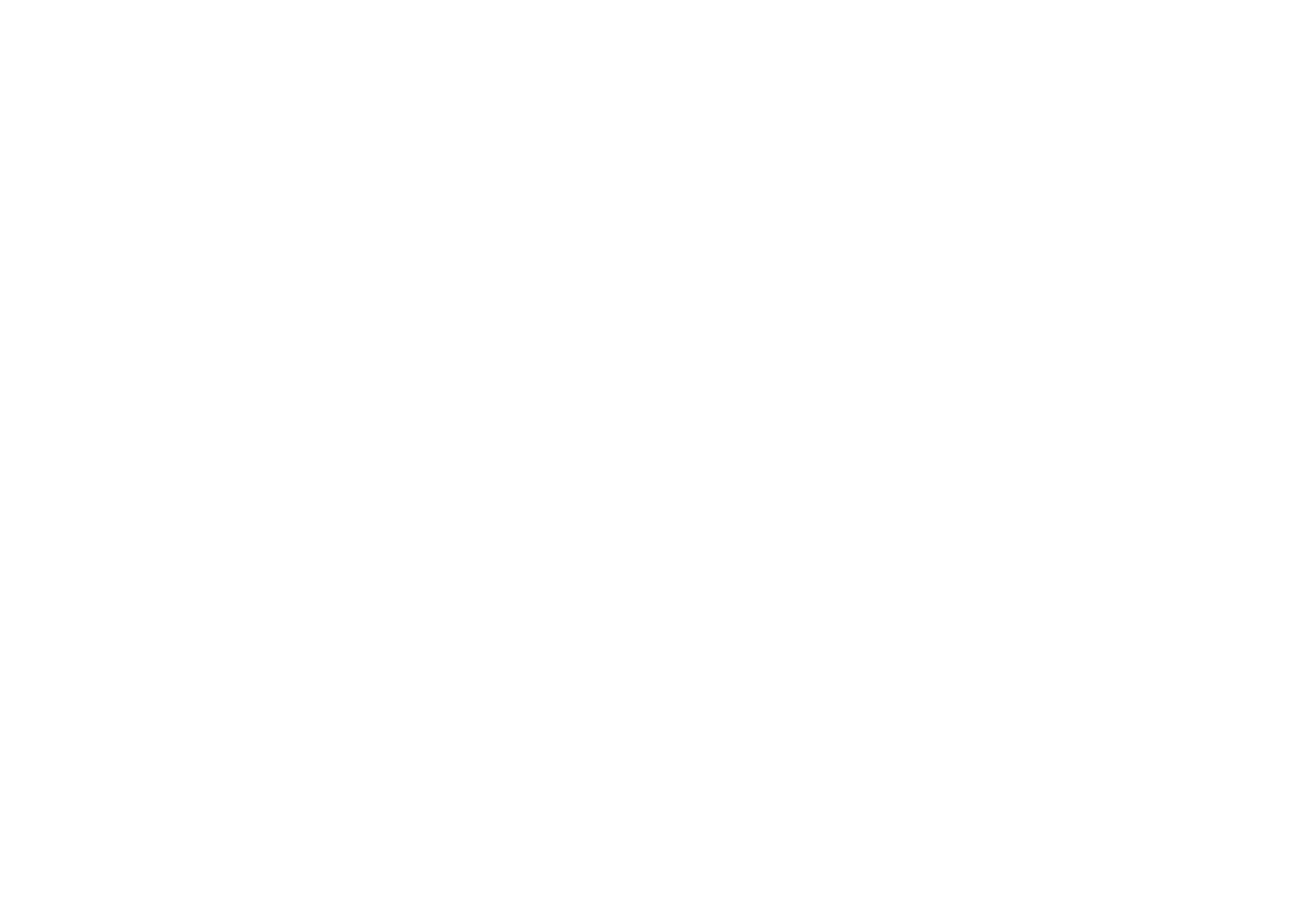}
\caption{Generalized light scattering susceptibilities of PCS and TFPI measurements of glycerol. High frequency data are taken from Ref.~\citenum{Brodin:2005a}. The normalization and fits are explained in the text.} 
\label{fig:glycerin-LS}
\end{figure}

Fig.~\ref{fig:glycerin-LS} shows selected autocorrelation functions of the electric field $g_1(t)$ from PCS, which were Fourier transformed to yield generalized susceptibility spectra between 190 and 260\,K, together with TFPI data from Brodin et al.\cite{Brodin:2005a} The TFPI data are in agreement with our own reference TFPI measurements with respect to peak position and spectral shape at a given temperature. We describe the depolarized light scattering spectra of glycerol by an $\alpha$-relaxation with high-frequency wing contribution by using an extended generalized gamma (GGE) distribution.\cite{Blochowicz:2003a}
For comparing PCS with BDS data, a reasonable normalization for the light scattering is needed. This is achieved by first normalizing the measured  TFPI spectral densities $S(\omega,T)$ with respect to the intensity of a Raman line. Then the generalized susceptibility is obtained by dividing by the Bose factor, which in the classical limit leads to \cite{bee:1988,callen:1951,hansen:1990}:
\begin{equation}
 \chi''(\omega) = \frac{1}{\hbar} \frac{S(\omega)}{n(\omega,T)+1} \approx \frac{\omega}{k_{\text{B}}T}\,S(\omega,T) 
\end{equation}
Here, the prefactor $1/T$ mimics a Curie law, which is well known behavior for the relaxation strength in dielectric spectroscopy. In order to combine TFPI with PCS data to produce a broadband light scattering data set, Fourier transformed correlation functions $g_1(t)$ are multiplied by $1/T$ and then the entire PCS data set is shifted in intensity relative to the TFPI data so that the high frequency wing of $\chi''(\omega)$ from PCS smoothly extrapolates to the low frequency flank of the susceptibility minimum observed in the GHz--THz range. The result of this procedure is shown in Fig.~\ref{fig:glycerin-LS}. We note that a single, temperature independent shift factor is applied for the whole data set. As both, the peak hight and the high frequency wing have to evolve in a continous fashion for all temperatures at the same time, the freedom in this procedure is rather limited and the uncertainty can be estimated to be around 10\%.

\begin{figure}[t]
\centering
\includegraphics[width=8.8cm]{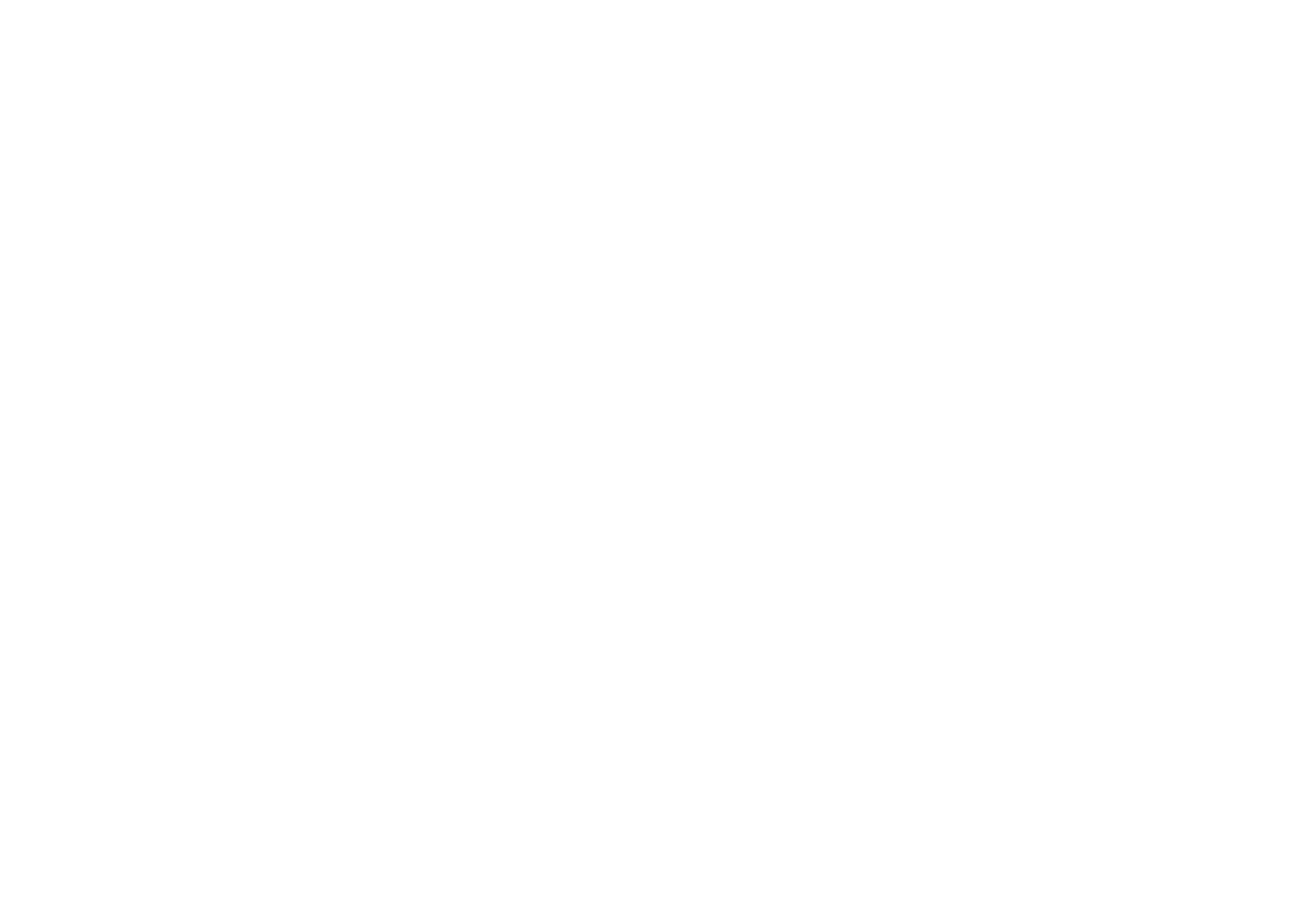}
\caption{Common fitting procedure for BDS and light scattering spectra of glycerol at 190\,K. The solid line shows the fit composed of self-  and cross-correlation contributions (dashed lines). High frequency data taken from \cite{Schneider:1998} at 185\,K and 195\,K } 
\label{fig:glycerin-LS-DS-normelized-190K}
\end{figure}
\begin{figure}[t]
\centering
\includegraphics[width=8.8cm]{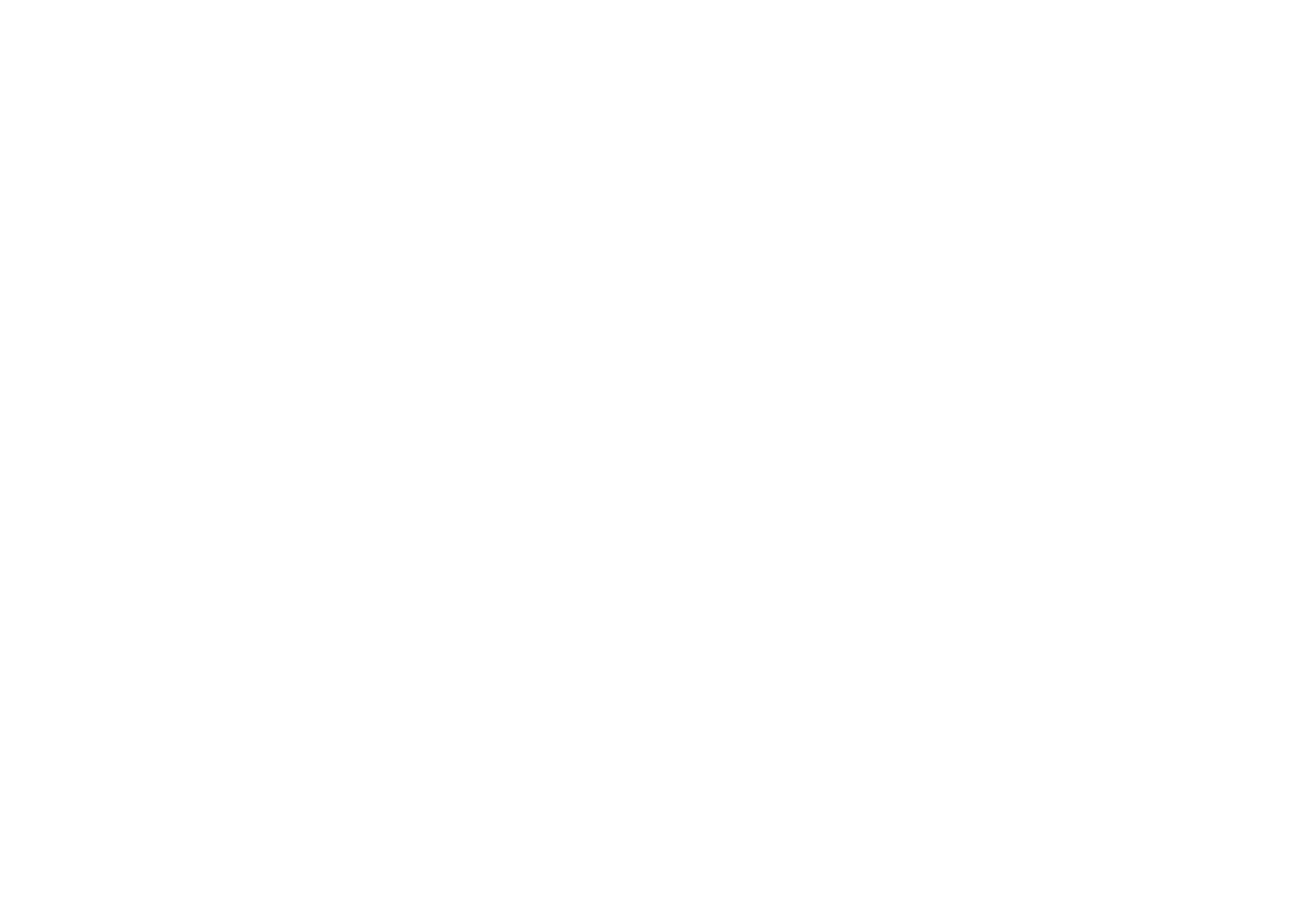}
\caption{Combination of dielectric and light scattering spectra of glycerol from Fig.~\ref{fig:glycerin-LS}. Dielectric permitivity spectra were composed of low frequency data from this work with high frequency data from Lunkenheimer et al.\cite{Schneider:1998} The light scattering data from Fig.~\ref{fig:glycerin-LS} were shifted with respect to the dielectric data with a common factor determined at 190K by overlapping the high-frequency wing of the $\alpha$-relaxation in both methods.} 
\label{fig:glycerin-LS-DS}
\end{figure}

Fig.~\ref{fig:glycerin-LS-DS-normelized-190K} shows glycerol data obtained by PCS, TFPI, Raman and BDS, all at $T=190\,$K. The light scattering data are normalized in intensity relative to the BDS data by superimposing the high-frequency wing of the $\alpha$-process in both methods. The reason for choosing this particular normalization is the observation in several monohydroxy alcohols that $\alpha$- and $\beta$-relaxation in both PCS and BDS perfectly superimpose and that this procedure allows to single out the Debye contribution in the dielectric spectrum of monohydroxy alcohols.\cite{Boehmer:2014a,Gabriel:2017a,Gabriel:2018a} The same idea is now applied for the polyalcohol glycerol, where such a superposition would allow to single out slow cross-correlation contributions in the dielectric spectra assuming that the self part of the correlation function is identical in both methods as in monohydroxy alcohols. The resulting combined light scattering and BDS data set, which shows the spectra of each method at the same temperature, is seen in Fig.~\ref{fig:glycerin-LS-DS}. We note that the scaling factor used to superimpose light scattering with the dielectric data is determined at the lowest temperature and is used for the entire data set at all temperatures. In the intermediate frequency range reasonable overlap is achieved up to the highest temperatures. We also mention, that in Fig.~\ref{fig:glycerin-LS-DS} all data, light scattering or dielectric, below 1\,GHz are from our own lab to ensure as far as possible identical temperatures, while the high-frequency dielectric data are taken from Schneider et al.\,\cite{Schneider:1998} and are slightly shifted in relaxation strength to match our own BDS results. The high frequency light scattering data are taken from Ref.~\citenum{Brodin:2005a} and merged with the PCS data set as described above.

Concerning the fits displayed in Figs.~\ref{fig:glycerin-LS-DS-normelized-190K} and \ref{fig:glycerin-LS-DS}, the PCS data are again described by the Fourier transform of a GGE distribution and the BDS data are described by the corresponding PCS fit at the respective temperature plus a Kohlrausch-William-Watts (KWW) stretched exponential function to take account of the BDS cross-correlation (c-c) part:
\begin{align}
  \Phi_\text{BDS}(t)=
  \Delta\varepsilon_{\alpha}\cdot \Phi_\text{PCS}(t) + 
  \Delta\varepsilon_\text{c-c}\cdot e^{-\left(\frac{t}{\tau_\text{c-c}}\right)^{\beta}}
  \label{eq:WW2}
\end{align}
with relaxation strength $\Delta\varepsilon_\text{c-c}$ and the stretching parameter being approximately $\beta\approx 0.85$ at all temperatures. Thus, the BDS spectra split into a broad self-correlation part and a more narrow cross-correlation part. Compared to monohydroxy alcohols the latter contribution is slightly more stretched, and it should be noted that in some monohydroxy alcohols the Debye-like relaxation also appears to be slightly broadened \cite{Arrese:2017}, in particular when the latter is close to the $\alpha$-relaxation indicating rather small supramolecular structures that comprise very few molecules only \cite{Gabriel:2018, Gabriel:2017a, Gabriel:2018a, Gabriel:2018c}. 

\begin{figure}[t]
\centering
\includegraphics[width=8cm]{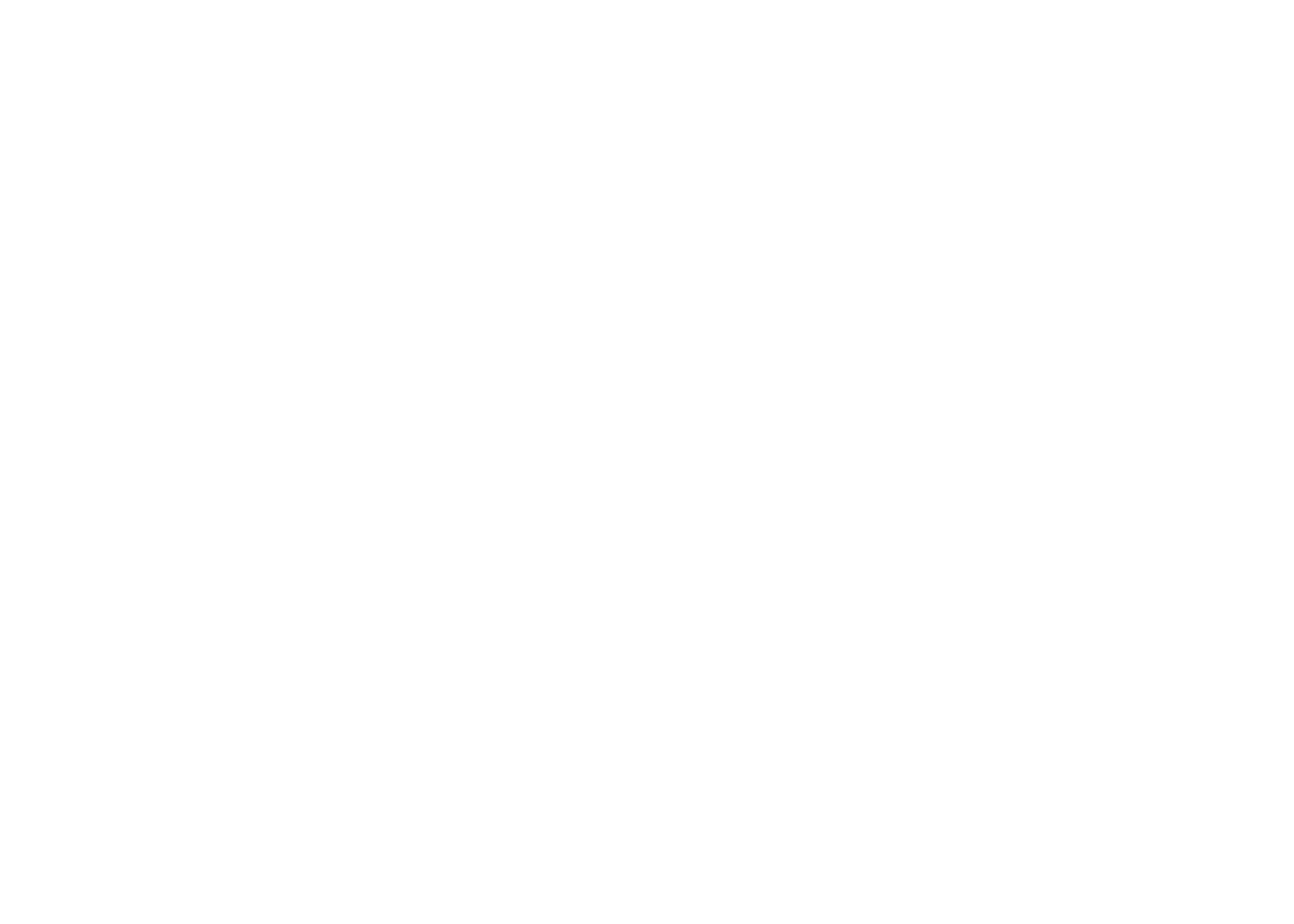}
\caption{Time constants for glycerol obtained by PCS, TFPI, Raman and BDS. The inset shows the temperature dependent Kirkwood correlation factor $g_k$ compared to the amount of cross-correlations approximated by the ratio of $(\Delta\epsilon_{\alpha}+\Delta\epsilon_\text{c-c})/\Delta\epsilon_{\alpha}$ obtained from the combined fits of the BDS data.} 
\label{fig:glycerin-tau-BDS-PCS}
\end{figure}

The PCS, TFPI, Raman and BDS time constants shown in Fig.~\ref{fig:glycerin-tau-BDS-PCS} are similar at high temperatures and separate slightly at low temperatures. This result is independent of the exact procedure of analysis, either fitting the dielectric loss independently with one GGE function or using the PCS fits plus an extra KWW for the dielectric data. We note that in the latter approach, the similarity of both timescales $\tau_\text{c-c}$ and $\tau_\alpha$ is consistent with the broadening of the cross-correlation part. In whatever way this cross-correlation contribution is interpreted in terms of physical mechanism, its close similarity with the $\alpha$-relaxation time scale indicates that to some extent dynamic heterogeneities are reflected in that process, in contrast to the case of most  monohydroxy alcohols, where larger objects, i.e.\ transient chains, are thought to average over the dynamically heterogeneous environment during reorientation.\cite{Gainaru:2010a}  

In order to crosscheck the applied procedure one can now use the relaxation strength of the self and the cross-correlation part in the spectrum and compare this with the Kirkwood correlation factor, which quantifies the amount of cross-correlations in a sample with interacting dipoles.\cite{Kirkwood:1939a,Froehlich:1958a} If the differences on the timescale of the microscopic dynamics are neglected, and further, if one assumes that the cross-correlations leading to $g_K>1$ are reflected dynamically mainly in the relaxation strength $\Delta\varepsilon_\text{c-c}$, then
a suitable measure for the amount of cross-correlations would be the ratio of the total relaxation strength over the $\alpha$-relaxation strength $(\Delta\epsilon_{\alpha}+\Delta\epsilon_\text{c-c})/\Delta\epsilon_{\alpha}$. The result is shown in the inset of Fig.~\ref{fig:glycerin-tau-BDS-PCS}. We note that the area under the peak of the microscopic dynamics in the light scattering spectra in Fig.\,\ref{fig:glycerin-LS} comprises about 10\% of the whole area under $\chi''(\omega)$. As this area is disregarded in the simple estimate for the cross-correlations in the spectrum, we include it in the determination of the indicated uncertainties. Now, this ratio can be compared with the Kirkwood/Fröhlich correlation factor:\cite{Kirkwood:1939a,Froehlich:1958a}
\begin{align}
g_\text{k}=\frac{9k_\text{B}\varepsilon_0 MT}{\rho N_\text{A} \mu^2}\frac{(\varepsilon_\text{s}-\varepsilon_\infty)(2\varepsilon_\text{s}+\varepsilon_\infty)}{\varepsilon_\text{s}(\varepsilon_\infty+2)^2}
\label{equ:kirkwood}
\end{align}
Here, $k_\text{B}$ is Boltzmann's factor, $N_\text{A}$ is Avogadro's number and $\varepsilon_0$ is the permittivity of the vacuum.
The static permittivity $\varepsilon_\text{s}$ was directly extracted from the dielectric data. Determination of $\varepsilon_\infty$, however, is not as straight forward. One way is to take $\varepsilon_\infty\approx 1.05\,n^2$ following the literature\cite{Dannhauser:1968b}, another approach is to take $\varepsilon'(\omega)$ at the highest frequencies reported in Ref.~\citenum{Schneider:1998} for one temperature $T=295\,$K and consider the temperature dependence by using the temperature dependence of the squared refractive index. Both approaches lead to slightly different values, which were used to determine the uncertainties for $g_K$. The refractive index $n$ and density $\rho$ were taken from Mirjana et al.\,\cite{Mirjana:2013} and Leron et al.\,\cite{leron2012densities} in the temperature range of $288-323\,$K and linearly extrapolated to lower temperatures. Finally a molecular dipole moment of $\mu=2.76\,$D \cite{Ryabov:2003} and a molecular mass for glycerol of $M=92.09\,$g/mole were used. 
The inset of Fig.~\ref{fig:glycerin-tau-BDS-PCS} demonstrates very good agreement of the values and the temperature dependence of both quantities in the range of $190-260\,$K, where a detailed comparison of the data is possible.


\section{Discussion}
\begin{figure}[t]
\centering
\includegraphics[width=8cm]{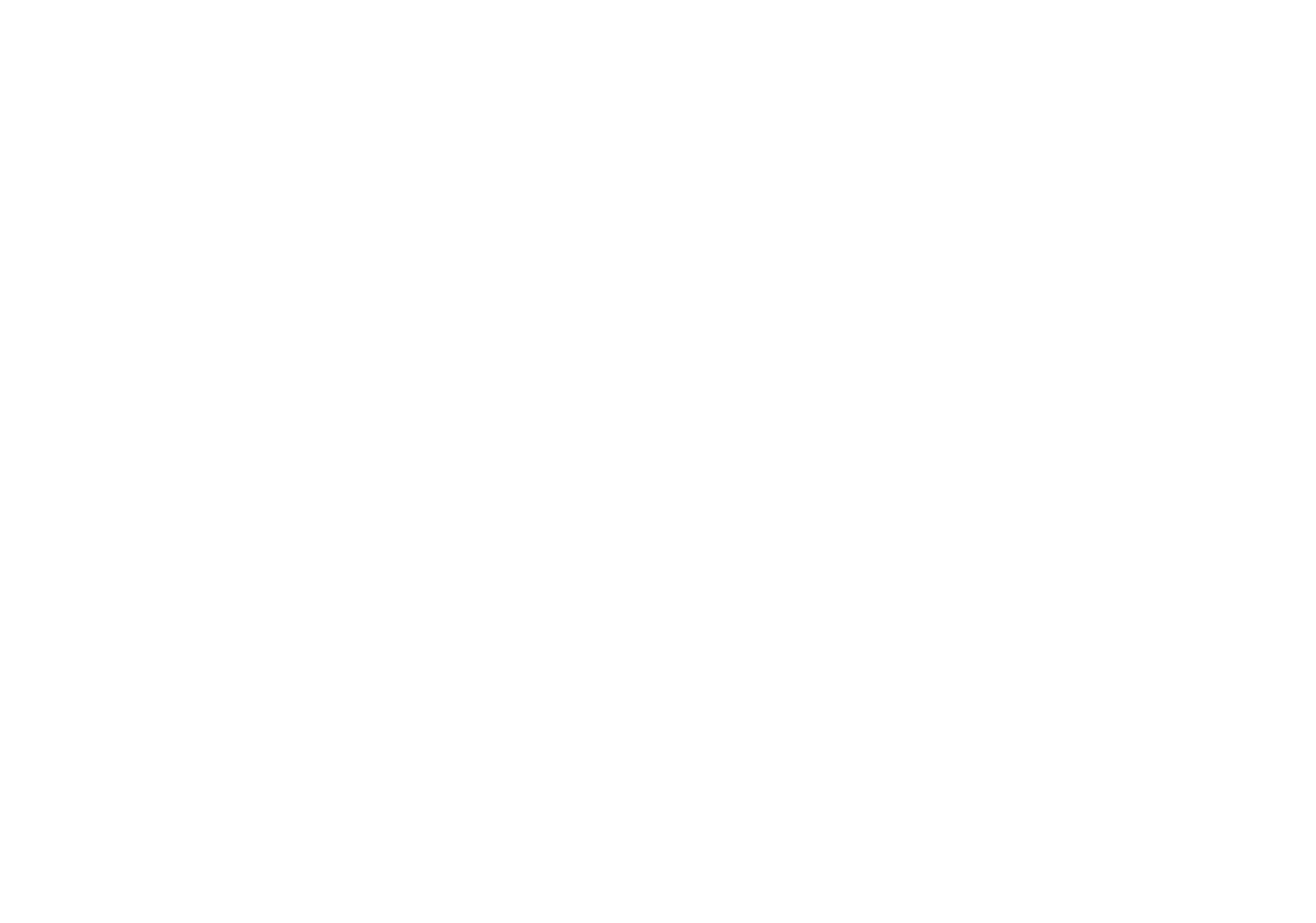}
\caption{A comparison of the generalized light scattering susceptibility from PCS with corresponding proton field cycling (FFC) data from Ref.~\citenum{Gainaru:2008} and field cycling data from a dilution experiment from Ref.~\citenum{Meier:2012}.} 
\label{fig:glycerol-FFC-PCS}
\end{figure}

In order to support the conclusions drawn from the above analysis in the following we compare our results with some literature data, namely fast field cycling NMR measurements  and dynamic shear relaxation spectroscopy. It is critical for our analysis, as we are aiming at separating self- from cross-correlation contributions in the dielectric spectra, that the  light scattering spectra are unaffected by cross-correlations.  Although this may not be true in general, as in some cases a corresponding Kirkwood correlation factor can also be measured in light scattering,\cite{Wang:2014a} in many systems indeed light scattering proves to be rather insensitive towards cross-correlations, in particular when the structural molecular anisotropy is not too big.  In the case of glycerol, this notion is further supported by NMR field cycling experiments: By measurements of frequency dependent spin-lattice relaxation times one can access the spectral density of an orientational self-correlation function based on the nuclear dipole-dipole interaction of protons.\cite{kruk2012} The field cycling technique in that case measures the rotation and translation of molecules due to the fact that there are intra- and inter-molecular proton-proton couplings. The masterplot of the generalized susceptibilities in Fig.~\ref{fig:glycerol-FFC-PCS} demonstrates that the PCS data are identical with the field cycling data taken from Gainaru et al.\,\cite{Gainaru:2008} in the region of the maximum in $\chi''(\omega)$ and at higher frequencies. The only difference is given by the translational contributions on the low frequency flank of the field cycling data originating from inter-molecular proton-proton interactions. In the case of proton NMR one can even replace some protonated glycerol molecules by deuterated glycerol and thus reduce the signal from the inter-molecular coupling by dilution as demonstrated by Meier et al.\,\cite{Meier:2012}. This removes the translational part in the spectrum and leaves only a pure orientational self-correlation. As both NMR and light scattering probe reorientation by a tensorial quantity, which leads to an $\ell=2$ correlation function in both cases, the observed identity in both shape and timescale demonstrates that indeed cross-correlational contributions are absent in light scattering spectra of glycerol in the region of the $\alpha$-relaxation. We note here, that of course correlation functions become different, when, e.g., in $^2$H-NMR experiments particular parts of the molecule are deuterated, like particular OH- and CH-groups as done by Döß et al.,\cite{Doess:2001} because then the dynamics of different molecular entities becomes distinguishable. By contrast, in the proton field cycling experiment as well as in light scattering, the involved interactions average over the whole molecule and thus produce the same spectral density in good approximation, see also supplement of Ref.~\citenum{Gainaru:2019a}.

Although it is clear in that way that light scattering reflects the self-part of the correlations, it is non-trivial that these should be the same in terms of spectral shape and timescale as the corresponding dielectric self correlation function. Even if one assumes that both to first approximation reflect the reorientation of the whole molecule, there is still the different rank Legendre polynomials involved in each correlation function. However,  spectral shapes and timescales which are identical within experimental limits are observed when comparing BDS and PCS data in the range of $\alpha$- and $\beta$-relaxation for many monohydroxy alcohols, which leads to the conclusion that the slow dynamics in monohydroxy alcohols is most likely governed by  processes that involve large reorientation angles as is the case, e.g., for random jump dynamics as compared to rotational diffusion.\cite{Gabriel:2018,Gabriel:2018a,Gabriel:2017a,Gabriel:2018c} Drawing on the observation in monohydroxy alcohols the same assumption was made in the context of the present analysis.

Interestingly, when light scattering and dielectric data are scaled on top of each other in the region of $\alpha$-process and high frequency wing as shown in Fig.~\ref{fig:glycerin-LS-DS}, then in the THz region the light scattering susceptibility shows about a factor three larger intensity than the dielectric loss.
In fact, this is exactly what is expected from general considerations for $\ell=1$ and $\ell=2$ correlation functions, if reorientational motions are restricted to very small angles. In that case it can be shown that:\cite{blochowicz:1999spectral, lebon:1997light, brodin:2003light}
\begin{equation}
  \chi''_{\ell=2}(\omega)\approx 3\cdot\chi''_{\ell=1}(\omega)
  \label{eq:smallangle}
\end{equation}
Previously it was argued that this relation can be applied for the dynamics in the region of the high-frequency wing, when the peak maximum of $\chi''_{\ell=2}(\omega)$ and $\chi''_{\ell=1}(\omega)$ are scaled on top of each other.\cite{Gainaru:2008,Flaemig:2019} If such scaling is applied, however, it can be estimated from Fig.~\ref{fig:glycerin-LS-DS} that the resulting difference in the THz dynamics, where the assumption of small angle restricted reorientation is most likely valid, would increase to about a factor of ten, inconsistent with Eq.~\ref{eq:smallangle}. By contrast, the scaling applied in Fig.~\ref{fig:glycerin-LS-DS} leads to an overall consistent picture up to the THz regime. 

In monohydroxy alcohols the appearance of the Debye contribution is related to the formation of supramolecular structures, which lead to cross-correlations of molecular dipole moments that produce an additional peak in $\varepsilon''$. By contrast, in polyalcohols the structures formed will most likely be network structures. How these are related to cross-correlations of molecular dipole moments is far less obvious. It should be noted however, that recently theoretical models suggest, that also without particular interactions like hydrogen bonds, simply the dipole-dipole interactions in a polar liquid may lead to cross-correlations that produce a separate slow relaxation peak in the dielectric loss, similar to the Debye process in monohydroxy alcohols \cite{Dejardin:2018}. Thus, the cross-correlations observed in the dielectric loss of glycerol need not necessarily be due to the hydrogen bonding network\,\cite{Pabst:2019a}. Further investigations of strongly dipolar but non-hydrogen bonding liquids will shed more light on this question.

\begin{figure}[t]
\centering
\includegraphics[width=8cm]{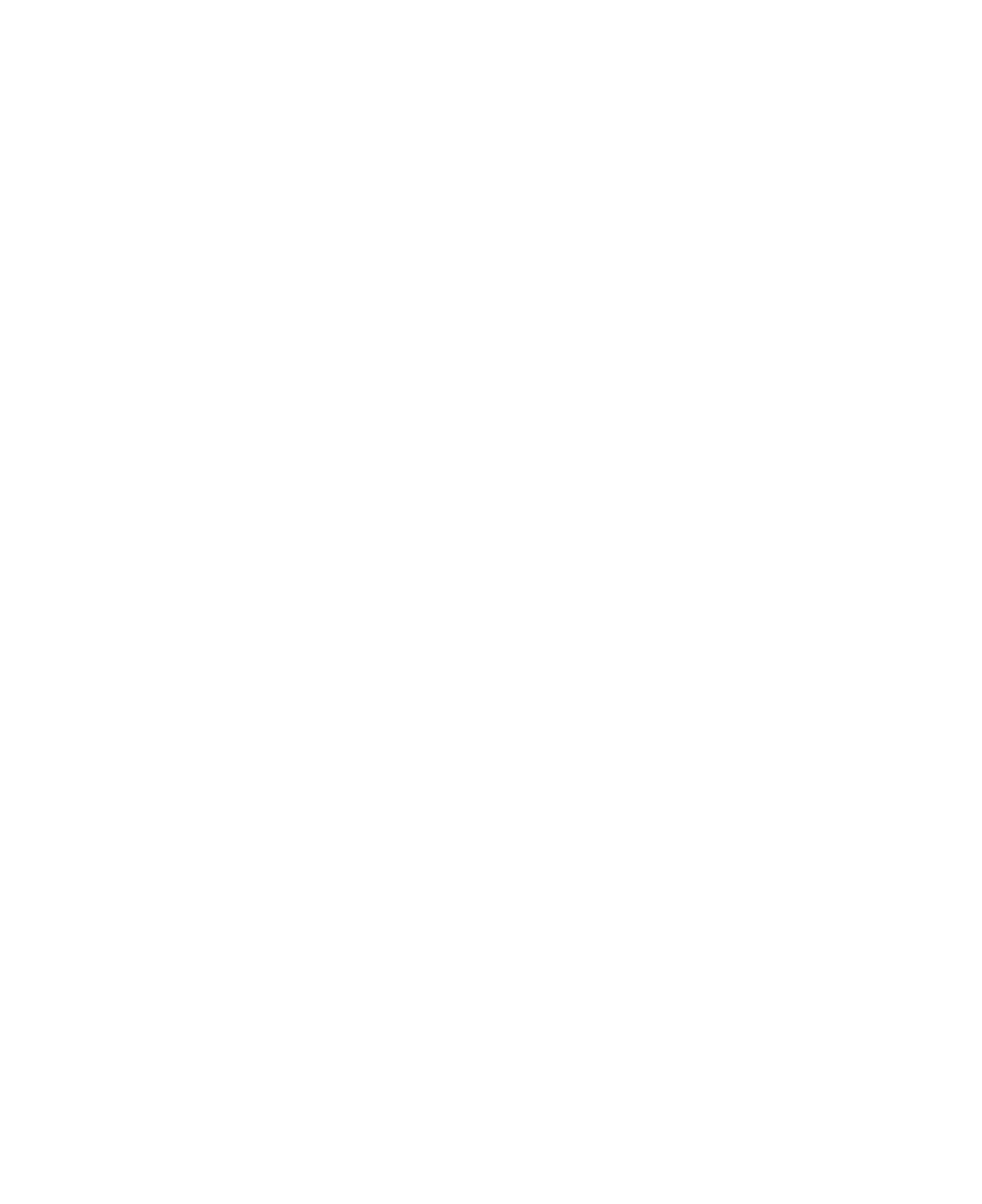}
\caption{PCS (blue), BDS (orange), and dynamic shear (purple) data of glycerol at 200\,K. The uper figure shows the modulus representation of BDS data from this work and dynamic shear modulus measurements from Ref.~\citenum{Jensen:2018}. Open symbols show a shifted spectrum for better comparison of shape. The lower figure shows a comparison of the light scattering susceptibility, the dielectric permittivity, and the relaxation part of the shear compliance (SC) without the fluidity contribution. The fit of the shear data (black line) is separated into a Debye (dasehd line) and a Cole-Davidson (dash-dotted line) contribution.}
\label{fig:glycerin-200K-modul-complience-rest}
\end{figure}
With the present analysis at hand, which suggests how to distinguish self- from cross-correlation contributions in the dielectric spectra, it is worthwhile to take a look at shear mechanical relaxation data, which by mere visual inspection clearly show a bimodal spectral shape in the case of glycerol. In Fig.~\ref{fig:glycerin-200K-modul-complience-rest} the spectral shape of the BDS measurements in modulus representation $M''(\omega)$ is compared to the dynamic shear modulus $G''(\omega)$ of glycerol from Jensen et al.\,\cite{Jensen:2018}. The comparison is made at 200\,K and both techniques have surprisingly similar spectral shape and deviate significantly from a simple peak structure. The difference in time constant might be due to a difference of $G_{\infty}$ and $M_{\infty} = 1/\varepsilon_\infty$ values, which change the time constants in modulus representation. In order to eliminate the influence of $G_{\infty}$ and $M_{\infty}$, we change the representation to a compliance representation $J^*(\omega)=J''(\omega)-i J(\omega)=1/G^*(\omega)$.\cite{Verney1989} Similar to the complex conductivity in dielectric spectroscopy $\hat\sigma(\omega) = i\omega\hat\varepsilon(\omega)$, where the low frequency limit $\hat\sigma(\omega\to 0) = \sigma_\text{DC}$ is often subtracted to reveal the relaxational contribution in the dielectric data, one can subtract the low-frequency limit of the complex fluidity $F^\ast(\omega)=i\omega\,J^\ast(\omega)$, i.e., $F^\ast(\omega\to 0)=F_0$, from the imaginary part of the compliance.
\begin{align}
J''_{\text{relax}}(\omega) = J''(\omega) - F_0/\omega
\label{equ:J}
\end{align}
Thus, $J''_\text{relax}(\omega)$ becomes directly comparable to the BDS permittivity and the light scattering susceptibility. In Fig.~\ref{fig:glycerin-200K-modul-complience-rest} the shear compliance data are described by a Cole-Davidson function with a shape parameter of $\beta_\text{CD} = 0.43$ and an additional slower Debye contribution and both functions approximately have the same intensity. For the shear data this result is independent of any amplitude scaling. Thus, Fig.~\ref{fig:glycerin-200K-modul-complience-rest} demonstrates, what can be identified as the  $\alpha$-relaxation is actually rather similar in all three experimental methods, while in a slower collective contribution each method seemingly shows very different aspects of the dynamics in glycerol: While light scattering pronounces the local aspect of the $\alpha$-relaxation on a molecular level, as cross-correlations most probably imposed by the H-bonding network do not play a significant role for the relaxation of the  optical anisotropy, for the dielectric relaxation cross-correlations are significant and also for the macroscopic shear response a slow collective relaxation mode seems to be important.

It is remarkable that like in glycerol, one can observe a slow contribution in the shear relaxation in nearly all monohydroxy alcohols\,\cite{Gainaru:2014a-PRL-098301}. By contrast, the self-correlation in PCS is free of any Debye-like contribution in glycerol, while for certain monohydroxy alcohols, especially secondary alcohols like 5-methyl-2-hexanol\,\cite{Gabriel:2018a}, a weak Debye-like contribution can be identified. The reason for this difference is still an open question. Most probably,  it arises depending on the average lifetime of the supramolecular structures and on how much restriction they impose on the reorientation probed by the $\alpha$-relaxation. If the restriction is significant and the $\alpha$-relaxation becomes slightly anisotropic, a small slow mode in the self-part of the correlation function may be seen, as is observed in light scattering, while otherwise the selfcorrelations are unaffected and a slow mode is only observed if cross-correlations play a significant role, like in dielectric spectroscopy. Further verification of this picture will be subject of future work.

\section{Conclusions}
A detailed comparison of the spectral shape of the dielectric and the light scattering response of glycerol in a broad spectral range, which involves only one temperature independent scaling parameter reveals that the dielectric loss can be approximately decomposed into a self- and a cross-correlation contribution.
The procedure is based on the observation that in the PCS spectra of glycerol cross correlation contributions are negligible to good approximation, as is demonstrated by a comparison of PCS with NMR relaxometry data. Second, a previous analysis of several monohydroxy alcohols showed that this self correlation appears in a very similiar if not identical fashion in both light scattering and dielectric spectra. Using the same analysis for glycerol  both contributions are harder to disentangle but still an overall consistent picture is achived, in which the Kirkwood correlation factor obtained from the static dielectric constant quantitatively aggrees with the relative strength of crosscorrelations obtained in the spectral decomposition and a factor of three in relaxation amplitude is recovered in the regime of microscopic dynamics, as expected for the different rank correlation functions in case of spatially restricted small-angle motions. Moreover, as we include shear relaxation data in the analysis, where a bimodal lineshape is directly obvious, it becomes clear that the data of all the different response functions of glycerol are compatible with the picture of an identical or at least very similar $\alpha$-relaxation, while every method shows a different expression of a slow collective relaxation mode, which is identified as a Debye peak in the case of monohydroxy alcohols.
Thus, surprisingly, the main difference between the various response functions of glycerol is found in the cross-correlation contribution. How far H-bonding plays the crucial role in establishing these cross-correlations or if electric dipole-dipole interactions in polar liquids are enough to produce such a slow collective process, as is suggested by a recent theory,\cite{Dejardin:2018} has to be left open for future studies.  

\section*{Conflicts of interest}
There are no conflicts to declare.

\section*{Acknowledgements}
The authors are grateful to Ernst Rössler, Bayreuth, for providing the original data from Ref.~\citenum{Brodin:2005a}, to Peter Schneider, Augsburg, for providing the data from Ref.~\citenum{Schneider:1998}, and to Tina Hecksher, Roskilde, for providing the original data from the Ref.~\citenum{Jensen:2018}. We are also grateful to Catalin Gainaru, Dortmund, for fruitful discussions and his suggestion to use Eq.~\ref{equ:J} when comparing dielectric permittivity with shear relaxation data. Furthermore, financial support by the Deutsche Forschungsgemeinschaft under Grant No. BL 1192/1 and 1192/3 is gratefully acknowledged. 

\bibliographystyle{achemso} 
\bibliography{glycerol}
   
\end{document}